\documentclass[12pt,tightenlines,nofootinbib,aip]{revtex4}
\usepackage{cjhebrew}
\usepackage{amssymb}
\usepackage{amsmath}
\usepackage{amsfonts}
\usepackage[stable]{footmisc}
\usepackage{hyperref}

\begin{document}

\title{In search of continuity: thoughts of an epistemic empiricist}

\author{Ian T. Durham}
\email[]{idurham@anselm.edu}
\affiliation{Department of Physics, Saint Anselm College, Manchester, NH 03102}
\date{\today}

\maketitle
\begin{center}
\noindent\emph{This essay is dedicated to the memory of my father-in-law \\ Lawrence Brod \textcjheb{h}$^{\prime\prime}$\textcjheb{`} \\ December 30, 1942 - February 6, 2011}
\end{center}
\section{Framing the debate}

What does it mean for something to be physically continuous?  Does it mean the object can't be broken down into individual parts or does it simply mean the individual parts are intrinsically linked?  And what about reality itself, particularly as it is manifest in the physical universe within which we abide?  These are heady questions that have long troubled physicists, mathematicians, philosophers, and even theologians.  As the great French physicist Louis de Broglie noted,
\begin{quote}
[t]he conflict between the continuous view in Physics, and its opposite, has existed through many centuries with varying fortunes, each gaining an advantage over the other in turn, and neither winning a definite victory.  For the philosopher there is nothing surprising in this, since the development of theory in every sphere of intellectual activity shows him that, if pushed to an extreme and opposed to each other, the concepts of both the continuous and the discontinuous are unable to give a correct rendering of Reality, which requires a subtle and almost indefinable fusion of the two terms of this antimony. \cite{Broglie:1939fk}, p. 217
\end{quote}
What is perhaps most intriguing about de Broglie's statement is the last line in which he suggests that a proper rendering of physical reality requires some `indefinable fusion' of the continuous and the discrete.  There are two points to be considered here.  The first is the most obvious: whether or not reality really requires \emph{both} continuous and discrete views as opposed to one in lieu of the other.  The second is more subtle: are we speaking of physical reality here or merely our \emph{rendering (knowledge) of it}?  It seems to me that these are two distinct questions to be considered for it is entirely conceivable that the continuity of our \emph{knowledge} about reality might be wholly distinct from the continuity of reality itself.

The distinction between states of reality and states of knowledge is described quite succinctly by Spekkens in \cite{Spekkens:2007fk} where he applies the terms \emph{ontic} and \emph{epistemic} to refer to states of reality and knowledge respectively.  In order to fully understand the distinction between these two types of states, let us look at simple, classical (and uncontroversial) examples of both.    The state of a clock, for instance, at a specific time at which we can specify the exact alignment of all the gears, hands, and dials, is considered an ontic state.  Conversely, the state of a molecule in a volume of gas, as specified, say, by its translational kinetic energy, is an epistemic state since it can only be predicted probabilistically using standard statistical mechanics.  As Spekkens points out, the distinction between these two types of state breaks down for states of \emph{complete} knowledge. So, for instance, if we could `freeze' our hypothetical gas at an instant in time and hone in on this one molecule, \emph{classical} physics assumes that we could specify its exact state (since it allows for the specification of instantaneous values for quantities such as velocity - I will have more to say on this a bit later). 

Spekkens' epistemic interpretation of quantum mechanics thus deals specifically with states of \emph{in}complete knowledge.  But what about those states of complete knowledge?  Are they anything more than a pipe dream?  Can we ever really know something fully?  As it turns out, these questions are intimately related to the debate between the continuous and discrete views.  Thus we can frame the continuity debate in terms of ontic and epistemic states.  Arguably we are most interested in exploring the continuity of the ontic states since the real point of all this is to understand reality.  But what if the only way to get information about ontic states is \emph{through} epistemic states?  Further, what if the epistemic states themselves are discrete?  How could we even determine if the underlying ontic states were continuous or not if the `lens' through which we view them is discrete?  These questions, however, deal with individual states rather than reality as a whole.  The process becomes considerably more murky when we ask questions about reality itself.  Is it even possible to reach conclusions about a reality that we are \emph{within} and thus an integral part of?

We may never find answers to some of these questions, but we can still learn a great deal about the universe and our ability to analyze it by considering them.  In doing so, however, we will find that the comfortable bosom of classical physics is nothing more than a mirage built on hundreds of years of human instinct and nature.  In the process we have so tightly woven mathematics and physics together that it sometimes becomes difficult to extricate one from the other.  Forgotten in this process is the fact that it is not at all clear just how closely mathematics comes to physical reality nor if it can serve as the sole conduit for interpreting that reality \cite{Durham:2010fk}.  Rightly or wrongly, we nevertheless often assume that most mathematics is representative of reality.  If the mathematics works, has experimental support, and, in the best of circumstances, predicts new phenomena, then it seems justifiable to take this view.  But what if two such mathematical models find an \emph{interpretive} contradiction?  Are we prepared to give up the more cherished interpretation?

\section{A paradox, a paradox, a most ingenious paradox\protect\footnote{A\lowercase{pologies to }G\lowercase{ilbert and }S\lowercase{ullivan}.}}
Classical, Newtonian physics was the \emph{only} physics for nearly three-hundred years.  It would not have developed in the manner and at the rate it did were it not for the development of `infinitesimal calculus' by Newton and Leibniz in the 1660s.  Though it has undergone a plethora of changes over the years, calculus remains an important tool in numerous fields.  Its development was not free from debate, however.  As its name suggests, one of its cornerstones was the notion of `infinitesimal change.'  In 1734 George Berkeley attacked the entire notion of these infinitesimal changes (termed `fluxions' by Newton).  He asked,
\begin{quote}
[a]nd what are these Fluxions? The Velocities of evanescent Increments? And what are these same evanescent Increments? They are neither finite Quantities nor Quantities infinitely small, nor yet nothing. May we not call them the Ghosts of departed Quantities? \cite{Berkeley:1734fk}, section XXXV.
\end{quote}
The ghosts to which Berkeley was referring can be understood through a simple example.  Suppose we have a simple parabolic function, $y=x^2$.  What happens to this function if we want to change it just a bit, say by an amount $\Delta y$?  This should mean that $x$ changes by some small amount as well, call it $\Delta x$.  The function should now look like this:
\begin{equation*}
y+\Delta y = (x + \Delta x)^{2}.
\end{equation*}
What's the actual nature of the small change, $\Delta y$, that we have just introduced?  Noting that our original function sets $y=x^2$, let's subtract it out from each side and simplify.  This gives us
\begin{equation*}
\Delta y = (x + \Delta x)^{2} - x^{2} = 2x\Delta x + (\Delta x)^{2}.
\end{equation*}
But how does introducing a small change in $x$ (essentially perturbing the system) affect $y$?  Taking the ratio of the changes gives
\begin{equation*}
\frac{\Delta y}{\Delta x}=2x + \Delta x.
\end{equation*}
This of course is \emph{not} the same thing as the derivative of $y$ which is known to be $2x$.  In order to get this familiar result it is necessary to simply assume that $\Delta x$ is \emph{infinitesimally small} and thus essentially zero \cite{Keisler:2010kx}.  But if it is essentially zero, what does that mean for the ratio on the left?  Either it is infinite (or undefined) or we need some more clever way out.  The clever way out didn't exist in 1734 and so Berkeley viewed this as a paradox: somehow the infinitesimal $\Delta x$ simply ``departed'' leaving its ``ghost'' behind in the denominator on the left.

There are numerous `resolutions' to Berkeley's paradox.  The foundations of infinitesimal calculus were put on a sound footing by Cauchy and Weierstrass in the early nineteenth century through the introduction of limits and a definition of continuity in terms of infinitesimals \cite{Strichartz:2000kx}.  Abraham Robinson later claimed to ``vindicate'' Leibniz (and presumably Newton) in his 1966 book \emph{Non-standard analysis} \cite{Robinson:1996uq}.  However, nearly every argument that purportedly gives infinitesimals a solid foundation shares the fact that they are all purely mathematical.  If we are solely interested in the validity of the mathematical argument then this is as it should be.  But what happens when we start applying infinitesimals to physical properties?

\section{Every good myth needs an Ancient Greek}
Actually, it is the converse question - how do we explain certain apparent physical paradoxes in a logical manner? - that helped lead to the development of infinitesimals to begin with.  The paradoxes to which we refer are those famously attributed to Zeno of Elea, notably those of Achilles and the tortoise, dichotomy, and an arrow in flight.  As Boyer writes, ``[i]f the paradoxes are thus stated in the precise mathematical terminology of continuous variables $\ldots$ the seeming contradictions resolve themselves'' \cite{Boyer:1959fk}, p. 295.  The importance of Zeno's arguments, however, lie not merely in the fact that they are seemingly easily refuted.  As philosopher John Mansley Robinson observes,
\begin{quote}
The art of Socrates is the art of Zeno put to a different purpose.  To the end it gave shape and form to the philosophy of Plato; and through Plato it entered into the very structure of European philosophy.  For it was Plato's pupil Aristotle who gave to logic the form which it was to bear for the next two thousand years; and to an extent which we are only beginning to realize this logic was a rationalization of the techniques of debate.  It was for this reason that the founders of modern science, when they rebelled against the logic of Aristotle, did so on the ground that it was concerned not with discovery but with proof.  Through Aristotle the spirit of Zeno passed into the thought of the late middle ages. \cite{Robinson:1968uq}, p. 139.
\end{quote}
For the record, I am no Aristotelian.  Nevertheless, it is in the sense described by Robinson that we can trace through Aristotle directly to Zeno one of the major problems that has crept into modern physics: an over-reliance on the use of mathematics as a means for \emph{interpreting} physical processes.  This problem is no more apparent than in the application of infinitesimals to the study of motion (which is precisely what Boyer was speaking of in regard to the resolution of Zeno's paradoxes).

Two of the most ubiquitous quantities in classical kinematics are velocity and acceleration, the latter being the derivative of the former and the second derivative of the position as a function of time.  Suppose a classical object has position $x_{1}$ at time $t_{1}$ and position $x_{2}$ at time $t_{2}$ and that this is the \emph{only} information we possess in relation to this object.  The most we can conclude from this is that the object had an \emph{average} velocity,
\begin{equation*}
\bar{v}_{\textrm{avg}}=\frac{\Delta \bar{x}}{\Delta t}=\frac{x_{2}-x_{1}}{t_{2}-t_{1}}\hat{x}.
\end{equation*}
We have no idea what happened to the object in the interim; its velocity could have varied considerably or not at all.  The arguments of infinitesimal calculus, viz. Cauchy and Weierstrass, are used to turn an average quantity such as this into an instantaneous one by invoking limits, e.g.
\begin{equation*}
\bar{v}=\lim_{\Delta t \to 0}\frac{x(t+\Delta t)-x(t)}{\Delta t}=\frac{d\bar{x}}{dt}.
\end{equation*}
Mathematically, there is nothing wrong with this.  But let us now consider this from the standpoint of empirical, i.e. physical, measurements.

\section{A speed trap\protect\footnote{T\lowercase{he author takes no responsibility for speeding tickets incurred by readers.}}: the discreteness of epistemic states}
First let us suppose our object is rigid which, by the usual rules of classical kinematics, allows us to treat it as a point particle localized at its center-of-mass (let's ignore any contact interactions).  Now suppose we wish to measure its speed with increasing accuracy.  While mathematics claims it is possible for us to measure it instantaneously, how might we accomplish this \emph{empirically}?  One of the most accurate ways to measure the speed of classical objects is with a radar (or speed) gun.  The way a radar gun works is that it bounces microwave signals off of a moving object and measures the Doppler shift of the signals.  As the object moves, the wavelength of the microwave signal changes and a phase shift arises between the wavelength measured at an instant, $t_{1}$, and a later instant, $t_{2}$.\footnote{This is technically true for both pulsed-Doppler and continuous-wave radar methods.}  The assumption is made that the electromagnetic wave is \emph{classical} and thus continuous.\footnote{Typically, `classical light' is that described by Maxwell's equations.}  But is classical light \emph{real}?  Couldn't classical light really be just a convenient approximation that relies on the the assumption that the photons in the beam are too numerous to effectively count?  Classical light supposedly is free of quantum behavior, but couldn't it be that the quantum behavior is `washed out' by the `noise' of additional photons, analogous to picking out a particular person from a crowd (the larger the crowd, the less likely it is that you'll find that person without some additional information)?  This latter proposal is similar to coarse-graining arguments in thermodynamic and quantum systems which have been used by Brukner and Zeilinger to argue that the continuum is nothing but a mathematical construct, a view I wholeheartedly endorse \cite{Brukner:2005vn}.  But even if we buy the classical light concept for now, what happens if we wish to pinpoint the location of our object's center-of-mass (unrealistically) to an \emph{exact point}?

We have two things to consider then: a) all light is really quantum, i.e. classical light is merely an approximation or b) classical light is real and truly distinct from quantum mechanical light.  In the first instance, it is clear why our attempt to find a truly instantaneous velocity will fail: since the wavelength of quantized light is proportional to its energy, in the limit as $\Delta t \to 0$ the time-energy uncertainty relation prevents us from determining the change in energy and thus the phase shift which, by extension, prevents us from measuring the velocity of the object.  Quantum systems have a limit to the accuracy with which they can be measured.  But this is well-known and only tends to force a retreat to the classical interpretation of light.

So what happens in the limit as $\Delta t \to 0$ for classical light?  One could argue that an instantaneous velocity, by definition, occurs at a specific location in space and yet is defined in terms of a \emph{change} in location, which makes no sense.  However, those who believe in the reality of instantaneous states will simply cite the infinitesimal calculus as `proof' of the instantaneous state.  While this smacks a bit of circularity, there is another argument against instantaneous velocity that can be made on more physical grounds.

Suppose we decrease $\Delta t$ while leaving $\Delta x$ unchanged.  As $\Delta t$ gets smaller and smaller, it implies we are measuring the difference between $x_{1}$ and $x_{2}$ more and more rapidly.  Lest we forget, classical physics limits how rapidly information can propagate.  At some point, without changing $\Delta x$, we will be \emph{empirically} prevented from further reducing $\Delta t$ since the ratio of $\Delta x$ to $\Delta t$ cannot exceed the speed of light.  So, if we wish to take $\Delta t \to 0$, we must take $\Delta x \to 0$ in order to keep the ratio at or below the speed of light.  But now we are faced with a bit of a problem.  The classical theory of light assumes light is a wave \emph{which is an inherently non-local phenomenon}.  In order to describe something as a wave, it is either necessary for it to have some spatial extent or for its overall function to be known so that local curvature can be deduced.  Clearly if we are forced to give it some spatial extent, then $\Delta x$ has a non-zero lower bound and the speed of light then forces $\Delta t$ to also have a non-zero lower bound which would seemingly prevent the measurement of an instantaneous velocity.  In other words, while the mathematical limit of $\Delta x/\Delta t$ as $t \to 0$ is $dx/dt$, this is \emph{not} necessarily true \emph{empirically}.

But, do we know the function?  Perhaps there is a way to measure a localized aspect of the wave.  Since it is an electromagnetic wave we know it is some sort of sine function so, in theory, we can calculate the so-called `local wavelength' (essentially a measure of a wave's localized curvature) by first finding its square \cite{Moore:2003fk},
\begin{equation*}
[\lambda(x)]^{2}=-\frac{4\pi^{2}f(x)}{f^{\prime\prime}(x)}
\end{equation*}
where $f(x)\propto \textrm{sin}(2\pi x/\lambda)$.  Recall, however, that we are using the Doppler effect in order to find the speed of our object.  The speed of the object produces a change in wavelength (which leads to a phase shift at the detector) in the incident wave that is directly proportional to the object's speed.  This means that $f(x)$ is dependent on the speed of the object and is thus \emph{not known ahead of time}.  Therefore we can't calculate the local curvature without already knowing the object's speed!  We're caught in a `Catch-22.'  Our only recourse is to conclude that it is impossible to measure a truly instantaneous velocity.

So regardless of whether we consider the light from our radar gun as being classical or quantum, there is a limit to how accurately it can measure velocity.  Now, any attempt to measure velocity (or just about any other physical quantity, for that matter) requires an interaction between the observer and the system under observation (sorry folks, there's just no way around that).  Any interaction necessarily requires an exchange of information.  In nearly every single classical physics experiment known to exist, electromagnetism - and thus light - plays a role, either directly or indirectly, in the process of making a measurement on the system.  Thus it seems that while it is clearly \emph{mathematically} possible for an instantaneous velocity to exist, we are \emph{physically} prevented from ever measuring one!  In this sense, it appears that states of complete knowledge, in which the values of system variables are known at every instant and in every point in space, are unobtainable.  This would seem to imply that epistemic states are ultimately discrete on some level: our \emph{knowledge} of the universe is discontinuous.

\section{Yet another Ancient Greek}
What is it that makes us cling to continuity?  The answer to that lies in the ``dim and distant past, as it were''\footnote{``\ldots in fact so long ago it was damn nearly \ldots in the year one.'' Or once so said Ian Anderson of the band Jethro Tull when introducing a live performance of the song `Skating Away on the Thin Ice of the New Day.'}.  As Trudeau points out, at the dawn of the 19th century, Euclid's \emph{Elements} were regarded as the ``supreme example of airtight deductive presentation'' \cite{Trudeau:2008fk}.  This apparently airtight system `sprung a leak' in the 19th century that led to the rich development of non-Euclidean geometries.  Nevertheless, Trudeau says, ``Euclid's text \ldots has been the scientific paradigm for most of scientific history'' \cite{Trudeau:2008fk}.  It was long thought that Euclid's axiomatic constructions \emph{required} the continuity of space.  One of the first to challenge this assumption was mathematician Richard Dedekind in 1893 \cite{Dedekind:1893uq}.  Dedekind and others it is possible to construct discontinuous spaces in which Euclidean geometry holds \cite{Strong:1898vn}.  Regardless, the association between Euclidean geometry and continuity was deeply ingrained in the scientific psyche for two millennia.  Given that geometric theorems were treated as objective truths that were knowable through intuition and/or reason and axioms were treated as obvious implications of definitions (see, for example, \cite{Bourbaki:1994ys}), it is then not a stretch to think that the prevailing view held that the universe \emph{itself} was continuous.  In fact it is doubtful, despite de Broglie's contention, that anyone prior to the twentieth century truly believed in a discontinuous universe, though they may have pondered the possibility.  Indeed, one could argue that this stubborn persistence of belief in (or perhaps even natural human `need' for) continuity is what led to (at least partially) the development of hidden variable theories.\footnote{Note that my argument in this essay does not necessarily preclude hidden variable theories, though neither does it imply them.  It is decidedly agnostic on the issue.}

Of course, while results from the Wilkinson Microwave Anisotropy Probe (WMAP) have demonstrated that the geometry of the universe must be flat to better than 1\% \cite{Spergel:2007fk} and thus `Euclidean,' we of course have long known that it is locally curved.  In general relativity, spacetime is normally modeled as a four-dimensional Lorentzian manifold (see, for example, \cite{Misner:1973uq}) on which one can measure a Lorentzian distance.  There is a direct relationship between causality on a given spacetime and the continuity of the Lorentzian distance on that spacetime \cite{Beem:1996kx,Minguzzi:2009vn}.  This idea simply formalizes the somewhat intuitive notion that causality is somehow related to continuity.  To get a better conceptual understanding of this, suppose two events, \emph{A} and \emph{B}, are causally connected.  Then there must be some way to get information from one to the other without exceeding the speed of light (or, more formally, they must be either \emph{timelike} or \emph{lightlike} separated).  If spacetime is \emph{dis}continuous, how do we know that this information couldn't `jump around' from point to point?  Continuity guarantees that the information follows a nice, orderly `path' between \emph{A} and \emph{B}.  This should make it easy to see the conceptual attraction of a continuous reality.  The problem is, how do we \emph{empirically} prove any of this?

Quantum field theory combines quantum mechanics with \emph{special} relativity and so \emph{technically} deals with a flat (Euclidean) spacetime.  Attempted extensions of quantum field theory to other manifolds have \emph{usually} retained the causality-continuity connection.  What quantum field theory actually does is quantize the \emph{field}.  This has the attraction of solving the problem of identical particles (i.e. technically any two electrons in the universe are indistinguishable to some extent) since any two particles are simply different `dimples' in the same field, so-to-speak.  By quantizing fields we have seemingly turned something inherently continuous and non-localized into something discrete and localized while somehow maintaining certain causal and continuous aspects of the non-quantized field.  To be clear, quantum electrodynamics, which is a quantum field theory, is the most accurate scientific theory ever developed, agreeing with experiment to within ten parts in a billion ($10^{-8}$) \cite{Particle-Data-Group:2010ys}.  But, ultimately, quantum field theory \emph{is built on quantum mechanics} just as classical field theory is naturally consistent with classical mechanics.  Thus, if there is a problem in either classical or quantum mechanics, as it appears there \emph{is} from our \emph{gedankenexperiment} with the radar gun, this problem extends to the associated field theories.  

We seem to be stuck.  Any continuity to the universe is implied and not measured since our ability to gain knowledge about the universe is necessarily discrete.  Our only other recourse, then, is to assume that mathematical `objects' have some kind of ontological status.  The problem with this view is that there is no way to \emph{prove} the ontological status of a mathematical object (one could always argue it is simply a representation of a physical object and is thus of a wholly different nature).  Nevertheless, this is an ongoing subject of debate (see for instance \cite{Durham:2010fk,Baker:2009uq}).

\section{Finis}
So, is the universe digital or analog?  Since it appears that our ability to gain \emph{knowledge} about the universe is discrete, we will likely never know unless we can find a clever way around the problem of determining the continuity of something through a discontinuous lens.  To my mind, quantum field theory is merely a way to `have our cake and eat it too.'  Certainly it has been wildly successful, but at heart it is still a quantum theory.  So perhaps the more enlightening question would be, are all `quantum' theories necessarily discrete?  The mere mention of the word `quantum' implies a `yes' answer to this, though perhaps the quantum field theorists would argue this point.  At any rate, it is my personal opinion that what makes quantum field theory so successful is its inherent `quantumness.'  Classical physics, with its inherent continuity, is nothing more than a convenient myth.  It's a nice approximation that works just fine when we don't look too closely.  To be sure, it gets a few things right and it does seem to be the natural way to explain gravity.  But this is simply because we don't need better results in these cases.  Gravity is so weak, we usually don't need to look too closely at it.  Perhaps the problems that arise when we do, result from the fact that we're not \emph{supposed} to look at it that closely; it's like trying to make out a van Gogh by looking at a single brush stroke.

Either way, there is clearly a limit to how much we can know about the universe regardless of whether it is classical or quantum (or both).  The fact that this limit even exists implies that our knowledge about the universe is necessarily discrete.  Even if Spekkens succeeds in developing an epistemic theory based on continuous variables, it would simply mean that individual epistemic states could be continuous.  The collection of \emph{all} epistemic states is necessarily more limited than the collection of all ontic states.  It is simply impossible to know everything there is to know about the universe (though it is not clear that this is necessary in order to definitively answer the question).  In addition, if Spekkens succeeds, his proof will likely be mathematical rather than empirical.  It is conceivable that a clever empiricist will find a way around this, but my guess is that the results would still be subject to interpretation.

\section{Coda}
When I was about two-thirds of the way through this essay my father-in-law passed away quite unexpectedly.  I suppose it is somehow fitting that this essay deals with continuity.  Larry was like a second father to me and I have chosen to dedicate this essay to his memory.  As I said in my eulogy to him, I've never been very good at saying goodbye.  Finality is not my fort\'{e}.  Yet I have reached a conclusion in this essay that contradicts my personal feelings.  That is how science is supposed to work.

The universe is a strange and wonderful place.  I would be disappointed if we could know everything there was to know about it.  Though the universe itself has cleverly prevented us from determining whether or not it is continuous, I'd like to believe that it is, at least in some respect, since then Larry could live on, even if it is just in an intangible way.  After all, the mere act of imagining something, since it is a part of us, means that it must be a part of the universe.  \begin{flushright}.\textcjheb{rb.h} $_{\prime}$\textcjheb{Mwl/s}\end{flushright}

\begin{acknowledgements}
I would like to thank my friend Barry Sanders for assisting me with the Hebrew that I included in the dedication.  \textcjheb{h}$^{\prime\prime}$\textcjheb{`} is an abbreviation of \textcjheb{Mwl/sh wyl`} and roughly translates to `on whom should be peace' when used after the name of someone who is deceased.  Apparently it is even occasionally used in a joking manner which would suit Larry just fine since he had a crazy sense of humor.  .\textcjheb{rb.h} $_{\prime}$\textcjheb{Mwl/s} (written right-to-left as in Hebrew) means `goodbye, friend.'
\end{acknowledgements}

\clearpage
\bibliographystyle{apsrev4-1}
\bibliography{FQXi2011Bib.bib}
\end{document}